\documentclass{llncs}

\usepackage[parfill]{parskip}
\usepackage{graphicx}
\usepackage{amssymb}
\usepackage{epstopdf}
\usepackage[T1]{fontenc}
\usepackage{caption}
\usepackage{subcaption}
\usepackage{amsmath}
\usepackage{proof}
\usepackage{bm}
\usepackage{listings}
\usepackage{fp}
\usepackage{xcolor}
\usepackage{colortbl}
\usepackage{booktabs}
\usepackage{multirow}
\usepackage{balance}
\usepackage{tabularx}

\usepackage{caption}
\usepackage{subcaption}
\usepackage{proof}
\usepackage{color}
\usepackage{amsmath}
\usepackage{amsfonts}
\usepackage[utf8]{inputenc}
\usepackage[english]{babel}
\usepackage{textcomp}
\usepackage{acronym}

\newcommand{\term}[2][]	{\ifthenelse{\equal{#1}{}}{\index{#2}}{\index{#1}}\emph{#2}}

\newcommand{\idx}[2][]	{\ifthenelse{\equal{#1}{}}{\index{#2}#2}{\index{#1}#2}}

\newcommand{\eaci}[1]{\edef\tmp{\noexpand\index{\ac{#1}}} \emph{\tmp}}

\newcommand{\defvoc}[2]{\expandafter\def\csname #1\endcsname{#2} {\newacro{#1}{#2}}}

\newcommand{\vf}[1]{\def\tmp{\index{\csname #1\endcsname@{\acf{#1}}}}\tmp\emph{\acf{#1}}}
\newcommand{\vs}[1]{\def\tmp{\index{\csname #1\endcsname@{\acf{#1}}}}\tmp\acs{#1}}
\newcommand{\vl}[1]{\def\tmp{\index{\csname #1\endcsname@{\acl{#1}}}}\tmp\acl{#1}}

\begin{document}
    
\title{Towards Assurance-Driven Architectural Decomposition of Software Systems}


\author{Ramy Shahin}
\institute{University of Toronto, Toronto, Canada\\
	\email{rshahin@cs.toronto.edu}}
\maketitle

\begin{abstract}
	Computer systems are so complex, so they are usually designed and analyzed in terms of layers of abstraction. Complexity is still a challenge facing logical reasoning tools that are used to find software design flaws and implementation bugs. Abstraction is also a common technique for scaling those tools to more complex systems. However, the abstractions used in the design phase of systems are in many cases different from those used for assurance.
	In this paper we argue that different software quality assurance techniques operate on different aspects of software systems. To facilitate assurance, and for a smooth integration of assurance tools into the Software Development Lifecycle (SDLC), we present a 4-dimensional meta-architecture that separates computational, coordination, and stateful software artifacts early on in the design stage. We enumerate some of the design and assurance challenges that can be addressed by this meta-architecture, and demonstrate it on the high-level design of a simple file system.
\end{abstract}

\keywords{Software Design, Assurance, Decomposition}

\section{Introduction}
\label{sec:intro}


Computer systems are so complex, so they are usually designed and analyzed in terms of layers of abstraction. An operating system typically runs on top of the bare hardware. A run-time system possibly comes next, and then the sets of abstractions introduced to programmers by a programming languages. Each layer can restrict the power of the layer underneath, but it can not be more powerful. For example machine instruction sets allow for arbitrary jumps to instruction addresses, while many high-level programming languages do not allow that.

Abstractions allow software designers to manage the inherent complexity of systems. Complexity is still a challenge facing logical reasoning tools that are used to find software design flaws and implementation bugs. Scalability (or lack thereof) of many such tools (e.g., model checkers) make them only suitable for relatively simple systems or toy examples. Abstraction, again, is a common technique for scaling those tools to more complex systems. However, abstractions used in system design are in many cases different from those used for assurance.

In this paper, we argue that the software abstractions used for assurance can also guide software designs, making them simpler, easier to understand, and readily suitable for automated reasoning. In particular, we present a \emph{4-dimensional software meta-architecture} that separates the computational, coordination, stateful, and meta-programming aspects of a software system early on in the design process. We argue that this systemic separation allows different assurance tools and techniques to be applied to each of the meta-architecture dimensions.

The rest of this paper starts with outlining some of the challenges of quality assurance of monolithic systems (Sec.~\ref{sec:problems}). The 4-dimensional meta-architecture is then presented in Sec.~\ref{sec:multicore}. This meta-architecture is then demonstrated on a simple file system design (Sec.~\ref{sec:example}). Related work is then briefly discussed in Sec.~\ref{sec:related}, and we finally conclude and outline some future directions in Sec.~\ref{sec:conclusion}. 
\section{Challenges of Assurance of Monolithic Systems}
\label{sec:problems}

In this paper, we refer to systems that do not separate computational, coordination, and state abstractions as \emph{monolithic systems}. This section outlines some of the challenges primarily caused by this lack of separation.
 



\subsection{Automated Logical Reasoning}


Integrating automated reasoning tools into the Software Development Life Cycle (SDLC) can significantly improve the quality of software systems by finding design/implementation flaws early in the process. Due to the inherent complexity of software systems, many automated reasoning techniques and tools can efficiently operate only on \emph{abstractions} rather than the concrete system artifacts. For example, model checkers operate on transition systems capturing abstract representations of system behavior. Theorem provers on the other hand are typically used to prove the correctness of abstract representations of algorithms.

Abstraction is a prevalent design technique allowing software designers to manage system complexity. However, design abstractions are in many cases different from the abstractions used in automated reasoning. For example, objects, classes, and interfaces are typical abstractions used in object-oriented design. On the other hand, when applying a software model checker to a program, program state is abstracted using bounded unrolling of loops. As a result, automated reasoning abstractions are \emph{synthetic} in many cases. One serious consequence is abstractions becoming out of sync with concrete system artifacts, leaking flaws into the system even when they are proven not to exist on abstractions. Decomposing system artifacts into different categories suitable for different reasoning and/or assurance techniques early in the design process provides a synergy between design and reasoning abstractions.

\subsection{Partial Correctness}


The axiomatic verification literature carefully, and rightfully, distinguishes Total Correctness from Partial Correctness. Total Correctness is accomplished when given a guarantee of the correctness of a precondition, an implementation is proven to satisfy a given postcondition. Partial Correctness on the other hand describes the same ternary relationship between a precondition, a postcondition and an implementation only if an implementation terminates. Proving whether an implementation will always terminate is undecidable, so Partial Correctness is a weaker guarantee than Total Correctness.

Separating terminating computations from non-terminating coordination artifacts allows for designing modeling languages of computational algorithms that always terminate. In his seminal book The Art of Computer Programming, Knuth characterizes the notion of an algorithm in terms of five attributes, the first of which is finiteness. In his definition of algorithms, Knuth explicitly states that "an algorithm must always terminate after a finite set of steps"~\cite{Knuth:1997}. Making termination explicit when modeling computational algorithms thus does not limit algorithm designers.

\subsection{Performance Assurance}

Performance analysis has always been an important aspect of software engineering. Theoretical asymptotic analysis of algorithm time and space complexities~\cite{Knuth:1997} has been an established field of Computer Science, but its practical counterpart has not been fully materialized yet. We typically use profilers to measure the execution time or memory consumption of a specific implementation under a specific workload, but we cannot do that statically. Profilers are analogous to dynamic type checkers, potentially reporting problems at run-time rather than compile-time, and potentially missing problems that are not covered by the workload used for analysis. 

Real-time systems in particular would benefit a lot from statically analyzing the performance of programs to make sure they meet their real-time constraints. Embedded systems with limited memory and processing resources would also benefit a lot from checking the performance characteristics of a system before being deployed. Power consumption has also been an important performance metric of programs running on handheld systems. Reusable library designers and users can leverage resource consumption contracts on modules to declaratively and soundly distinguish high-performance components from other components meeting the same functional requirements but consuming more processing time, memory and/or power.

But automatic performance analysis is theoretically impossible in general, simply because almost all performance analysis problems reduce to the halting problem, which is undecidable. Several heuristic algorithms have been proposed to address the problems of termination (e.g.~\cite{Cook:2011}) and resource consumption. However, those are heuristics that can not be used in a logically sound and complete analysis.


Modeling algorithms using only terminating constructs would enable new performance analysis scenarios:
\begin{itemize}
	
\item Time Analysis: With algorithm models terminating by design, the compiler becomes responsible of making sure recursion is bounded and loops all have an upper bound on the number of iterations. Since the compiler needs either to infer those bounds or to require the model to make them explicit, then worst case asymptotic complexity can be directly calculated.
Asymptotic time complexity can be also added to the contract of an algorithm. The implementation will be checked against that contract, and clients using that algorithm will be "taxed" that upper bound on time complexity when their individual complexities are calculated. Those contracts can then be used by performance analysis tools to find performance bottlenecks statically.

\item Space Analysis: Similar to time complexity, space complexity can be also calculated directly. This will be an asymptotic analysis as well because algorithms usually abstract away platform-specific details for portability. Still, given those asymptotic bounds, lower-level compilation phases can come up with more accurate space analyses as they generate platform-specific code.

\item Power Analysis: Power consumption is at least as important as time and space in handheld systems. Time complexity, individual instructions, using specific peripheral devices and network bandwidth are among the factors affecting power consumption. Since we can at least asymptotically quantify each of those factors, we can again statically calculate an asymptotic bound on power consumption given performance contracts.

\item Bandwidth Analysis: Similar to Power, network bandwidth analysis can be performed based on time analyses, and the different levels of overhead added at different layers of a communication stack.
\end{itemize}



\section{A 4-Dimensional Meta-Architecture}
\label{sec:multicore}

\begin{figure*}[t]
	\centering
	\includegraphics[width=0.75\textwidth]{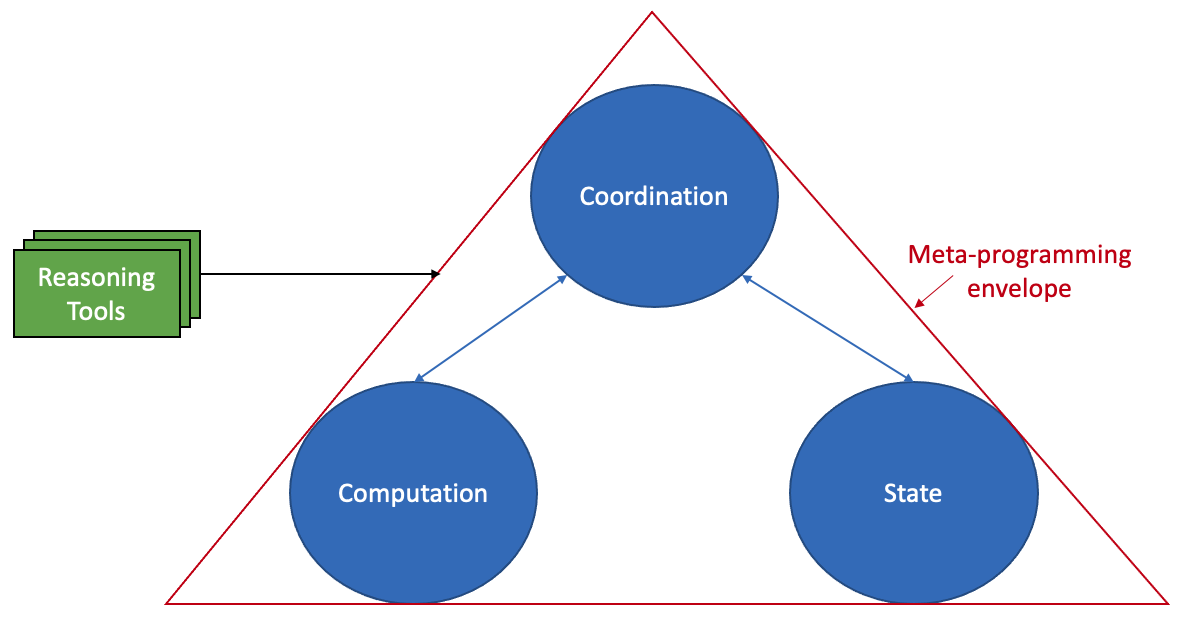}
	\caption{A 4-dimensional meta-architecture (architectural pattern) decomposing architectural artifacts into four models: coordination, computation, state, and meta-programming artifacts.}
	\vspace{-0.2in}
	\label{fig:multicore}
\end{figure*}


Given the differences in the nature of different software artifacts, we propose the meta-architecture in Fig.~\ref{fig:multicore}. This is a meta-architecture in the sense that it is system-independent, and can be instantiated for different systems. This is also sometimes referred to as an \emph{architectural pattern}~\cite{Bass:2012}. The three orthogonal models we identify here are \emph{computation}, \emph{coordination}, and \emph{state}. The coordination model interfaces with the others, accessing only constructs publicly exported. For example, a message handler in the coordination model might need to perform a computation. This can be achieved by calling a function exported from the computation model. Similarly, a transition from one object state to another might involve calling a computational function. In such case, the function is called by a process in the coordination model, and the result is used to atomically update the stateful object. 

To allow for direct integration with program reasoning tools, the exported constructs of all three models form a meta-programming envelope. Those constructs need to export programmable interfaces for models to integrate with each other, and also to be used by tools (e.g., verification, test-case generation, performance analysis). The three models presented earlier, in addition to the meta-programming envelope, form a \emph{4-dimensional meta-architecture} that can guide the design of software systems.

\begin{table*}[tbp]
	\caption{Examples of aspects of the different models within the 4-dimensional meta-architecture.}
	\label{tbl:multicore}
	\centering
	\scriptsize
	\resizebox{\textwidth}{!}{\begin{tabular}{llll}
	\toprule
			& Computation & Coordination & State \\
	\midrule
	Underlying Logic & Hoare logic, Constructive logics & Linear/Temporal logics & Description logics \\
	Calculus & $\lambda$-calculus & Process calculi (e.g., $\pi$-calculus) & Relational calculus \\
	Semantics & Denotational & Operational & Axiomatic \\
	Type systems & System F and dependent types & Session types & Type-state \\
	Verification & Contract-based & Model checking & Symbolic model checking \\
	\bottomrule
	\end{tabular}}
\vspace{-0.2in}
\end{table*}

The main advantage of splitting the architecture into multiple models is liberating software designers to use the formalisms that best suite the responsibilities of each of the models, instead of having to stick to only one set of formalisms throughout the design of the whole system. Table~\ref{tbl:multicore} presents a taxonomy of formalisms, with examples of particular formalisms that might be more suitable for particular models than others.

Software reasoning tools are usually based on an underlying logic.
For computation, Hoare logic~\cite{Hoare:1969} has been widely used to reason about sequential programs, and truth judgments are usually defined in constructive logics. Coordination on the other hand is more about causality and timing constraints. Temporal logics~\cite{Pnueli:1977} and Linear logics~\cite{Girard:1987} are capable of expressing those judgments. Description logics~\cite{Baader:2003} have been commonly used to formalize data representation, and thus might be best suited for stateful object models. 

Similarly, different kinds of formal calculi are applicable to the different modeling languages. Variants of $\lambda$-calculus (e.g., $\lambda$C, $\lambda$P, $\lambda$D~\cite{Nederpelt:2014}) have been designed with computation as their primary focus. Process calculi~\cite{Hoare:1978,Milner:1980,Bergstra:1989,Milner:1999} are all about modeling concurrent processes, communication channels and interactions. Relational calculus and Relational algebra~\cite{Date:2004} have been used for decades as the underlying formalisms for relational data management in databases. Many relational concepts are applicable to in-memory stateful objects.

Defining the semantics of language constructs is what gives models meaning. Different approaches to language semantics have been used over the years~\cite{Nielson:2007}. Denotational semantics model language constructs using mathematical functions, so they are a natural fit for computational models. Operational semantics model the operational effects resulting from the evaluation of constructs. Coordination is effectful, and we naturally tend to think about coordinating systems operationally. Axiomatic semantics focuses on defining logical invariants that are to be maintained across evaluations. This is exactly what stateful objects and their integrity invariants and constraints are to be defined on top.

Typed languages usually integrate their type systems together with their formal calculi. System F of polymorphic types~\cite{Pierce:2002} and dependent types~\cite{Nederpelt:2014} associates types (with varying expressive powers) with expressions. Concurrent systems need a different sort of type systems (e.g., Session types~\cite{Caires:2014}). Stateful objects are themselves treated as types in many language paradigms. Elaborate Typestate-based systems have been designed though to track the dynamic interfaces (types) of objects subject to logical object state~\cite{Strom:1986,Deline:2004,Aldrich:2009}.

There are several approaches to program verification, and the assurance of program properties in general. Again, different approaches would fit better than others to different models. Correctness of computations can be verified based contracts (pre-conditions and post-conditions). 
Model checking approaches are best-suited to verifying temporal properties of systems defined in process calculi~\cite{Tiu:2005,Bradfield:2018,Bunte:2019}. With invariants being first-class constructs of stateful objects, symbolic model checking~\cite{Burch:1992,Clarke:1996} can be used to verify that object state transitions do not violate object invariants.

\section{Example: A Simple File System}
\label{sec:example}

\begin{figure*}[t]

\includegraphics[width=\textwidth]{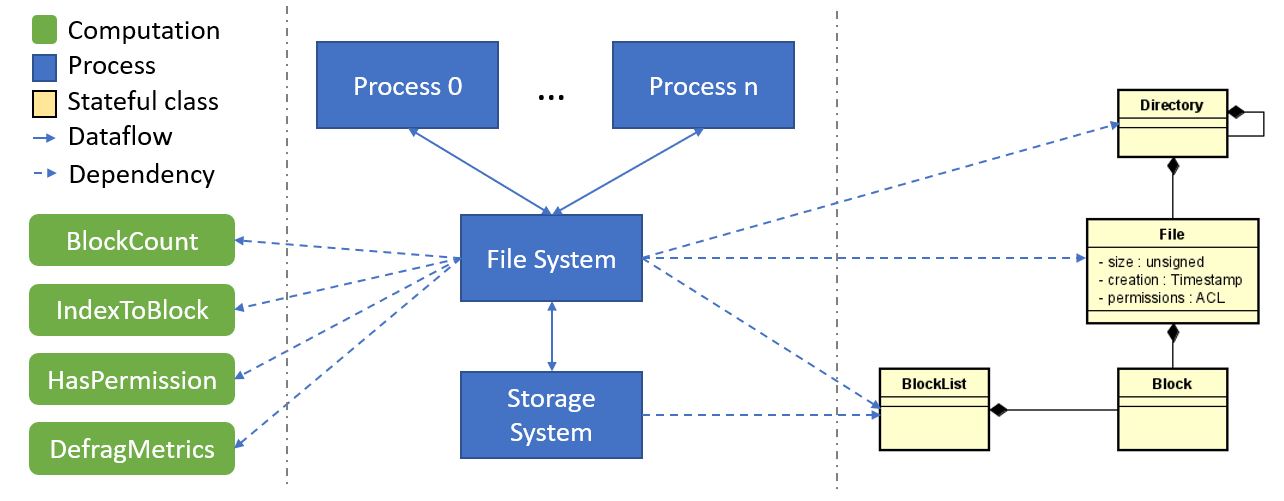}

\caption{A simple file system example. Diagram is split into three sub-models: stateful classes/objects (yellow), processes (blue), and computations (green).}
\label{fig:example}	
\vspace{-0.2in}
\end{figure*}

To demonstrate the concepts presented in this paper, we use a simple file system as an example (Fig.~\ref{fig:example}). A file system can be thought of as a process communicating with other processes. In addition to the user-mode processes that need to read and write data from/to files, a file system also interacts with a storage system that manages a block-based storage device (e.g., disk or tape), and possibly with other operating system processes. Different processes communicate with each other through message passing. Processes and their communication channels are modeled as blue model elements in Fig.~\ref{fig:example}. Their dependencies on stateful objects and computations are modeled as dashed arrows. A process can call a computational function, and can also query or update the state of an object. Computations and stateful objects do not directly interact though.

The stateful objects managed by the file system are modeled as a UML class diagram in Fig.~\ref{fig:example} (yellow model elements). Stateful objects include files, directories, and a list of storage blocks that can be allocated to different files. A file exclusively contains the set of blocks where its contents are stored. A directory contains a set of files and possibly sub-directories. Stateful objects have to preserve a set of invariants. For example, a block is either free, or belongs to strictly one file. Also the size of a file is a function of the number of blocks it contains. Structurally, the directory structure has to be acyclic, with each file/directory having at most one parent (i.e., directories form a tree).

In addition to communication between processes, and state transitions of stateful objects, a file system needs to compute the values that are to be passed across processes, or used to determine which state an object should transition to. Computations are modeled as green elements in Fig.~\ref{fig:example}. For example, when a file is created or appended, the number of blocks needed has to be computed using the \emph{BlockCount} function. Files do not necessarily occupy contiguous blocks on a storage device, so given a per-file block table, accessing a particular byte within a file, the index of that byte needs to be translated into a pair of values: a block identifier, and an index within that block. Those are computed using the \emph{IndexToBlock} function. The Access Control List (ACL) of a file encodes access permissions to that file, and determining whether a user, a group or a process has access to that file typically involves a computation (\emph{HasPermission}). Whenever possible, storing file contents in contiguous blocks improves access time due to locality patterns, especially for sequential file access. Defragmentation is a process where file contents are moved to unused blocks that are physically closer to other blocks used by the file. Deciding whether to defragment a volume is usually subject to several metrics that also need to be computed (\emph{DefragMetrics}).
	
File systems, pretty much like other operating system components, are usually cited as examples of non-terminating software systems. A file system has to continuously respond to requests from user processes. This is typically modeled as an \emph{event loop}, where a system waits indefinitely for an external event, and when that event arrives it is processed by the system, which then goes back to the waiting state. Read/write requests are examples of events processed by a file system. Modeling this event loop as a process rather than a computation makes it easier to assure the safety and correctness of the system. Coordination properties of processes (e.g., deadlock/livelock freedom) can be checked using a model checker without having to include computational states in the model. This can highly reduce the state-space of the model, improving scalability of existing model checkers. At the same time, contract-based assurance techniques, or axiomatic tools based on Hoare-logic, can be used to check the correctness of sequential computations without having to take the inherent concurrency of the system into consideration.

\section{Related Work}
\label{sec:related}

Systematic decomposition of software systems into smaller units has been the driving force behind several software engineering paradigms. Seminal work by Parnas~\cite{Parnas:1972} suggests hiding each design decision in a separate module. 
Decomposing systems statically into functions~\cite{Abelson:1996}, or objects~\cite{Booch:2004}, or deployment-time services~\cite{Erl:2005}, are among the most commonly used paradigms. Hybrid decomposition techniques have been also suggested~\cite{Chang:2001}. In addition, cross-cutting concerns inspired multi-dimensional decomposition techniques~\cite{Tarr:1999}, such as Aspect-Oriented Programming (AOP)~\cite{Kiczales:1997}, and Feature-Oriented Programming (FOP)~\cite{Prehofer:1997}.

The aforementioned techniques and paradigms base their decomposition decisions upon either problem domain abstractions, or encapsulation of design decisions. This paper on the other hand suggests an orthogonal dimension of decomposition, taking assurance techniques and their abstractions into consideration. This can be thought of as a generalization of multi-dimensional separation of concerns~\cite{Tarr:1999}, adding an explicit assurability dimension.

Separation of computation and coordination aspects of software systems has been argued for by Gelernter back in the early 1990s~\cite{Gelernter:1992}. In this paper we follow that argument, and it is one of the inspirations behind the 4-dimensional meta-architecture. It is unfortunate though that almost 30 years later, monolithic system architectures are still the norm.

\section{Conclusion and Future Work}
\label{sec:conclusion}

In this paper we argued that different software quality assurance techniques operate on different aspects of software systems. To facilitate assurance, and for a smooth integration of assurance tools into the Software Development Lifecycle (SDLC), we presented a 4-dimensional meta-architecture that separates computational, coordination, and stateful software artifacts early on in the design stage. We enumerated some of the challenges that can be addressed by this meta-architecture, and demonstrated it on a simple file system design.

For future work, we plan to study the adequacy of existing software modeling tools, and potentially provide tool support for the 4-dimensional meta-architecture. Integrating modeling with logical reasoning tools (e.g., model checkers, theorem provers, SMT solvers), and effectively combining the results computed by reasoning tools are two future research directions as well. Tooling support might involve the design of notations/languages suitable for the different aspects of the meta-architecture. Integration of results from different reasoning tools would involve proving that this integration preserves soundness.

\section* {Acknowledgments}
The author thanks the anonymous reviewers for their feedback and insightful suggestions.

\bibliographystyle{splncs04}
\bibliography{pl,datalog,se} 
\end{document}